\documentclass[aps, prl, reprint]{revtex4-1}

	\usepackage{amsmath}
	\usepackage{amsfonts}
    \usepackage{graphicx}

\newcommand{\p}{\partial}
\newcommand{\lang}{\left\langle}
\newcommand{\rang}{\right\rangle}

\DeclareMathAlphabet{\bi}{OML}{cmm}{b}{it}
\newcommand{\rmd}{\mathrm{d}}

\begin{document}

\title[Runaway electrons in stellarators]{Runaway electrons during a coil quench in stellarators}
\author{Pavel Aleynikov, Per Helander and H\aa kan M Smith}
\affiliation{Max-Planck-Institut f\"ur Plasmaphysik, 17491 Greifswald, Germany}

\begin{abstract}

It is shown that avalanches of runaway electrons can arise in stellarators, even if there is no net toroidal current in the plasma or the magnetic-field coils, if the current in the latter varies rapidly enough, e.g. due to a superconductor quench. In present-day devices such as W7-X, significant runaway generation is theoretically possible only at very low gas or plasma density in the vacuum vessel, e.g. between discharges, if a seed population of free electrons is present. In reactor-scale stellarators, more dangerous runaway generation may occur both between discharges and during low-density plasma operation. Since a radiation-induced seed population is necessarily present in an activated device, an accidental coil ramp-down could convert substantial magnetic energy into wall-damaging runaway currents. There is however much more time to mitigate such events than in tokamaks disruptions. 

\end{abstract}


\maketitle

\section{Introduction}

Tokamak disruptions are often accompanied by the generation of
``runaway'' electrons, which are accelerated to relativistic energies by the toroidal electric
field that is induced when the plasma current decays
\cite{Breizman,Helander-2002}. Runaway electrons have the potential of causing great damage and remain a serious safety concern for large tokamaks \cite{Salewski_2025}. They have occasionally been observed 
in stellarators as a result of Ohmic heating \cite{Bernstein}
and during ramp-up or ramp-down of the helical coil currents
\cite{ATF,TJ-II,Uragan}. In a superconducting stellarator like W7-X, the coil
currents are normally varied very slowly, but an exception occurs if
the superconductor suffers a quench or if the quench-detection system accidentally shuts down the coil currents for other reasons. In this case, the
current is ramped down on a time scale of a few seconds
\cite{Birus,RISSE2019910}. Although no net toroidal current is then interrupted
(since none of the coils extend around the torus in the toroidal
direction), a toroidal electric field is nevertheless induced, because
the coils create a poloidal magnetic flux, which decays during the
quench. It is thus possible that runaway electrons could be created,
which is the subject of the present Letter, where we discuss the generation, cross-field drift, and danger associated with such electrons.

\section{Magnetic and electric fields}
The magnetic field can be written in Boozer coordinates as \cite{ROP}
\begin{equation*}
\bi{B} = \nabla \psi \times \nabla \theta + \nabla \varphi \times
\nabla \chi,
\end{equation*}
where $\theta$ and $\varphi$ are the poloidal and toroidal angle
coordinates, $2 \pi \psi$ the toroidal magnetic flux and $2 \pi \chi$
the poloidal flux. These fluxes are functions of space and time, but
if the current in all coils quench at the same rate and the plasma
current is neglected, the geometry of the magnetic field remains
constant, so that
\begin{eqnarray*}
\psi(\bi{r}, t) = f(t) \psi_0(\bi{r}), \\
\chi(\bi{r}, t) = f(t) \chi_0(\bi{r}),
\end{eqnarray*}
and the Boozer angles are then time-independent. For simplicity, we
restrict our attention to the case of a low-$\beta$ plasma without net
toroidal current, so that most of the magnetic field is produced by
the coils. A suitable, single-valued, vector potential is
\begin{equation*}
  \bi{A} = \psi \nabla \theta - \chi \nabla \varphi,
\end{equation*}
and the electric field induced by the current quench is
\begin{equation*}
\bi{E} = - \p \bi{A} / \p t = - \dot f(t) (\psi_0 \nabla \theta - \chi_0 \nabla \varphi). 
\end{equation*}
It is important to note that the gauge matters: $\bi{E}$ is not
invariant under the transformation $\chi \to \chi + \rm const$. To
determine the correct value of the additive constant, we consider the
somewhat idealized case where the coil currents are replaced by
surface currents on some toroidal surface $S$ surrounding the
plasma. According to Faraday's law, the loop voltage
\begin{equation*}
  \oint_C \bi{E} \cdot \rmd\bi{r}
\end{equation*}
around a closed contour $C \subset S$ is equal to minus the time
derivative of the linked magnetic flux. If we choose $C$ so to
encircle the torus once toroidally, but not poloidally, then
\begin{equation*}
  \oint_C \bi{E} \cdot \rmd\bi{r}  = 2 \pi \dot f(t) \chi_0 = 0,
\end{equation*}
and we thus conclude that the additive constant should be chosen so
that $\chi$ vanishes on $S$. The function $\chi(\psi)$ then equals the
poloidal flux between the surface $S$ and the flux surface labelled by
$\psi$.

On the magnetic axis, where the enclosed toroidal flux vanishes, $\psi
= 0$, the field becomes $\bi{E} = \dot\chi \nabla \varphi$ and the loop voltage
\begin{equation*}
  U = \oint \bi{E} \cdot \rmd\bi{r} = 2 \pi \dot \chi.
\end{equation*}
As already noted, the right-hand side represents the time derivative of
the poloidal flux, which can also be calculated from the mutual
inductance $M_i$ between the coils ($i=1,\ldots 5$) and the magnetic
axis
\begin{equation*}
  U = \sum_i M_i N \dot I_i,
\end{equation*}
where $I_i$ denotes the current in the $i$'th coil and $N$ the number
of windings. In W7-X, the total
poloidal flux is typically about
\begin{equation*}
  2 \pi \chi = \sum_i M_i N I_i \simeq 2 \; \rm Tm^2
\end{equation*}
at full field. If the ramp-down of coil currents occurs on the time
scale of a few seconds, we thus expect a loop voltage of a few volts
and an electric field,
\begin{equation} E = \frac{U}{2 \pi R} = \frac{ \dot \chi}{R}, \label{E}
\end{equation}
of the order of 0.02 V/m. This field will quickly diffuse through the
vacuum vessel, whose skin time is only a few ms. For the same reason, very little of the
magnetic energy will be dissipated Ohmically in the vacuum vessel,
making this energy instead available to accelerate electrons.

\section{Acceleration and deconfinement of electrons}

Such electrons would not be of any great concern if they were
sufficiently well confined by the magnetic field. Since runaway
electrons have such small pitch angles that they are not trapped in
local magnetic wells, their orbits deviate very little from the
magnetic field, if this is kept constant \footnote{The radial orbit
excursion is of the order of $\Delta r \sim \gamma m_e c/(eB)$, which
is about 1 cm for a 10 MeV electron in W7-X. Here $\gamma$ denotes the
Lorentz factor, $m_e$ the electron rest mass, and $c$ the speed of
light.}. However, in a decaying magnetic field, the electrons will move
toward the wall because of the $\bi{E} \times \bi{B}$ drift. In a
vacuum magnetic field, $\bi{B} = G \nabla \varphi$,
\begin{equation*}
\frac{\bi{E} \times \bi{B}}{B^2} \cdot \nabla \psi = \frac{G (\nabla \varphi \times \nabla \psi) \cdot \bi{E}}{B^2}
= - \frac{\p \psi}{\p t},
\end{equation*}
implying that $\psi$ is a constant of the motion,
\begin{equation*}
\frac{\p \psi}{\p t} + \left( \frac{v_\| \bi{B}}{B} + \frac{\bi{E}
  \times \bi{B}}{B^2} \right) \cdot \nabla \psi = 0.
\end{equation*}
As the field decays, the electrons on each magnetic surface will thus move radially outward in such a way that the enclosed toroidal flux remains constant. Roughly speaking, if the field strength drops by a factor of two, their distance from the magnetic axis will grow by a factor $\sqrt{2}$. All free electrons will thus eventually leave the confinement region as $B \rightarrow 0$.  

This analysis is only strictly valid in the limit of small plasma pressure and current, and will cease to hold if so many electrons are accelerated that the resulting plasma current substantially modifies the magnetic field. Qualitatively, however, the $\mathbf{E} \times \mathbf{B}$ drift induced by the decaying coil currents will still convect particles outward. 

The characteristic time required for accelerating an electron to relativistic energies is
\begin{equation}
  t_\mathrm{c} = \frac{m_e c}{e E} \sim 100 \; \rm ms \label{tc}
\end{equation}
for $E=0.02$ V/m. This is much shorter than the quench time, so we conclude that practically all runaway electrons will reach MeV energies. The key question is whether such freely accelerating electrons will multiply through impact ionization and thus set off an avalanche. 

\section{Runaway avalanche}


{\it Avalanche in a neutral gas.} --- In both a plasma and a neutral gas, the friction force $F_s$ experienced by a charged particle decreases over a broad energy range. Consequently, electrons run away if their energy exceeds a threshold at which the accelerating force exceeds friction, $eE > F_{s}$.

\begin{figure}
    \centering
    \includegraphics[width=\linewidth]{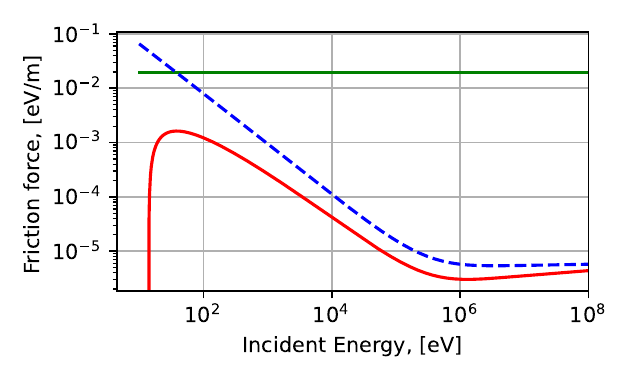}
    \caption{Friction force Eq.~(\ref{Bethe}) acting on an electron in a neutral hydrogen with with atomic density of $ n_\mathrm{e} \sim 4.8 \cdot 10^{15} \; {\rm m}^{-3}$  (red) and in a fully ionized plasma (dashed) with the same density. The green horizontal line indicates the maximum loop voltage expected during W7-X coil quenches.}
    \label{fig1}
\end{figure}

The friction on an electron in a neutral gas was derived by Bethe \cite{Bethe},
\begin{eqnarray}
  F_s = \frac{4 \pi r_0^2 n_e m_ec^2}{\beta^2} \Big( \ln \frac{m_ec^2(\gamma-1)\sqrt{\gamma+1}}{\sqrt{2} I} - \nonumber \\ 
   - \left(\frac{2}{\gamma} - \frac{1}{\gamma^2}\right)\ln{2} + \frac{1}{\gamma^2} + \frac{(\gamma-1)^2}{8\gamma^2} \Big),\label{Bethe}
\end{eqnarray}
where $r_0$ is the classical electron radius, $I$ the characteristic ionization potential, $n_e$ the total (bound) electron density, $\beta \equiv v/c$, and $\gamma = (1-\beta^2)^{-1/2}$. Equation (\ref{Bethe}) has a maximum near $\gamma_\mathrm{max} \approx 2.72 {I}/(m_ec^2 )+ 1$, where the friction force is
\begin{equation*}
  F^{\mathrm{max}}_s = \frac{2 \pi r_0^2 n_e m_e^2c^4}{2.72 I }.
\end{equation*}
Between discharges in W7-X, the pressure is about $10^{-5}$~Pa and the hydrogen molecule density thus $ n_\mathrm{n} \sim 2.4 \cdot 10^{15} \; {\rm m}^{-3}$ ($n_e = 2 n_\mathrm{n}$). Accordingly, all free electrons will run away if the electric field exceeds $E_\mathrm{max} \approx 1.5 \mathrm{\ mV/m}$, which is much smaller than the field induced during a quench. The force given by Eq.~(\ref{Bethe}) is shown in Fig.~\ref{fig1}. 

The particle energy loss described by Eq.~(\ref{Bethe}) arises primarily from ionization processes. Secondary electrons produced in these collisions may themselves be accelerated by the electric field, leading to a runaway avalanche \cite{Sokolov, Gurevich}. Although the ionization cross section decreases with increasing energy, the total number of secondary electrons generated during the acceleration to non-relativistic speeds is negligible compared to that produced by a relativistic electron \cite{Jayakumar}, as shown in the Appendix.

In the regime where all particles run away, the avalanche rate is simply given by the ultra-relativistic electron ionization rate. The Bethe cross section for the total ionization in the high-energy limit has the form \cite{kim2000extension} 
\begin{equation*}
  \sigma_i = \frac{4 \pi r_0^2}{\alpha^2 \beta^2} \left(M^2 \left( \ln(\gamma^2 -1)  - \beta^2 \right) + C \right),
\end{equation*}
where $M^2$ and $C$ are constants related to the continuum oscillator strength and $\alpha$ is the fine structure constant. For atomic hydrogen $M^2 = Q/2$ and $C = 2-Q - M^2 \ln \alpha^2$ with $Q=0.5668$. The resulting avalanche rate when $\beta \rightarrow 1$ is
\begin{equation}
  \Gamma_n = \frac{4 \pi r_0^2 n_e c}{2} \frac{m_ec^2}{I} \frac{Q}{2} \left( \ln\frac{m_e c^2(\gamma^2 -1)}{2 I} + \frac{4}{Q} - 3 \right), \label{Gamma_n}
\end{equation}
which represents the maximum possible rate. Taking $ n_e = 2 n_\mathrm{n} \sim 4.8 \cdot 10^{15} \; {\rm m}^{-3}$ and the logarithm to be equal to 15, we obtain $\Gamma_n \approx 15 \mathrm{\ s}^{-1}$, resulting in a net amplification 
    $$\frac{N_r(t)}{N_r(0)} = \int_0^{t} \Gamma_n dt, $$ 
which amounts to several tens of orders of magnitude within a few seconds when $eE > F_s^\mathrm{max}$. Significant runaway growth via avalanching is thus plausible in W7-X.

When the electric field decreases below $E_\mathrm{max}$, only electrons born with sufficiently high energy can run away. The minimum friction force, which occurs near $\gamma \approx 10$ (see Eq.~(\ref{Bethe})), defines the critical electric field $E_\mathrm{c} = \min(F_s)/e$ below which no acceleration is possible \cite{Connor},
\begin{equation*}
  e E_\mathrm{c} = 4 \pi r_0^2 n_e m_ec^2 \ln \Lambda,
\end{equation*}
where $\ln \Lambda$ is the first logarithmic term in Eq.~(\ref{Bethe}).
This definition is equivalent to the classical threshold electric field for ``runaway breakdown'' in a fully ionized plasma, provided that the Coulomb logarithm $\ln \Lambda$ is adapted to describe free-free collisions. To do so, the ionisation potential $I$ in Eq.~(\ref{Bethe}) needs to be replaced by $\hbar \omega_p \gamma$, where $\omega_p$ is the plasma frequency \cite{Breizman}.

For electric fields such that $ E_{max}> E > E_\mathrm{c}$, only particles born with energy above the critical energy
\begin{equation*}
\varepsilon_c \approx \frac{1}{2}\frac{E_\mathrm{c}}{E} m_e c^2
\end{equation*}
can run away. The rate at which runaway electrons are produced in a tokamak in this regime was calculated by Rosenbluth and Putvinskii \cite{Rosenbluth}. As shown in the Appendix, their calculation extends to the stellarator case with unimportant modifications and results in the growth rate
\begin{equation}
  \Gamma \simeq \sqrt{\frac{\epsilon \pi}{3 (Z+5)}} \frac{E/E_\mathrm{c}-1}{\tau \ln \Lambda}, \label{gamma}
\end{equation}
where the relativistic collision time is $\tau ={m_e c}/({e E_\mathrm{c}}) = (4 \pi r_0^2 n_e c \ln \Lambda)^{-1}$, the factor $\epsilon^{1/2}$ accounts for the fraction of trapped particle, and $Z$ is the effective ion charge experienced by relativistic electrons.

In a weakly ionized plasma, Eq.~(\ref{gamma}) should be used with caution, since collisions of secondary particles with neutrals are less efficient than collisions in a fully ionized plasma. The avalanche rate in a neutral gas is therefore expected to be higher. Refs.~\cite{Martin-Solis,Hesslow} suggested a modification to Eq.~(\ref{gamma}) that approximately accounts for the effect of reduced collisions. The suggested recipe amounts to using the critical runaway energy $\varepsilon_c$ as the characteristic energy when evaluating $\ln \Lambda$ in the case of a neutral gas.

\begin{figure}
    \centering
    \includegraphics[width=\linewidth]{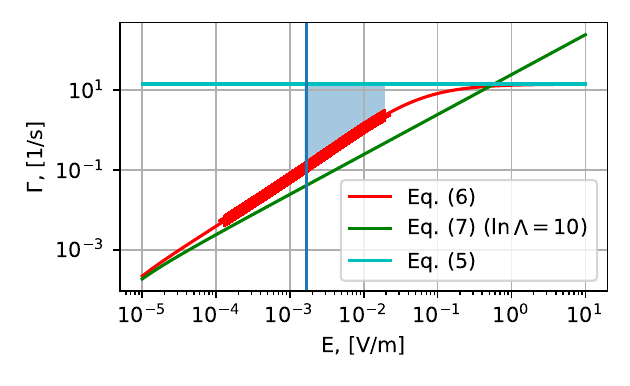}
    \caption{Avalanche rate as a function of electric field in neutral hydrogen (same density as in Fig.~\ref{fig1}). The vertical line indicates $E_\mathrm{max} \equiv F_s^{\mathrm{max}}/e$.}
    \label{fig2}
\end{figure}

Equations (\ref{Gamma_n}) and (\ref{gamma}) can be combined into an interpolation formula (assuming $E/E_c \gg 1$) between the two avalanche regimes, 
\begin{equation}
  \Gamma_c \simeq 4 \pi r_0^2 n_e c \left(\frac{1}{a}\frac{E_c}{E} + \frac{1}{q}\frac{I}{m_ec^2}\right)^{-1}, \label{Gamma_comb}
\end{equation}
where  $\ln \Lambda$ in $E_c$ is the logarithmic term in the parentheses of Eq.~(\ref{Bethe}) with $\gamma$ evaluated from the force-balance relation $eE = F_s$. Furthermore, $a = \sqrt{\epsilon \pi/{(3 (Z+5))}}$ and 
\begin{equation*}
q = \frac{Q}{4} \left( \ln\frac{m_e c^2(\gamma^2 -1)}{2 I} + \frac{4}{Q} - 3 \right) \approx 2.5,
\end{equation*}
taking $\gamma \approx 10$ for primary electrons.
Figure~\ref{fig2} shows the resulting avalanche rate as a function of the electric field, taking $Z=1$ and $\epsilon = 1$ \footnote{Note that the trapped particle fraction can be significant in stellarators, even on the magnetic axis. In W7-X, the trapped fraction on axis typically ranges from 0.15 to 0.3.}.

Taking $E(t) = E_0 e^{-t/t_{\rm quench}}$ with $E_0 t_{\rm quench} = \chi_0/R$, and $t_{\rm quench} = 3$ s (see \cite{RISSE2019910}) we obtain ${N_\mathrm{r}(\infty)}/{N_\mathrm{r}(0)} \approx 400$,
suggesting that a small seed population cannot give rise to a substantial runaway current. It should however be emphasized that Eq.(\ref{Gamma_comb}) is merely an interpolation, with the electric field of interest falling between rigorously described limits (indicated by a thick red line in Fig.~\ref{fig2}). This estimate assumes that only a small fraction of secondary electrons become runaways even when $E>eF_s^\mathrm{max}$ (shadowed region in Fig.~\ref{fig2}). Using instead Eq.~(\ref{Gamma_n}) for $E > eF_s^\mathrm{max}$ and Eq.~(\ref{gamma}) for $E < eF_s^\mathrm{max}$ results in an amplification factor ${N_\mathrm{r}(\infty)}/{N_\mathrm{r}(0)} \approx 10^{45}$, which represents the most ``pessimistic'' limit. The possibility of a strong runaway avalanche therefore cannot definitely be ruled out. More accurate kinetic modeling is required to refine the quantitative prediction.

{\it Runaway breakdown and plasma operation.} --- Equation~(\ref{Gamma_comb}) is valid in a neutral gas, but as runaway breakdown progresses and the gas is ionized, the secondary electrons experience more efficient collisions with the ionized plasma (see Fig.~\ref{fig1}). Therefore, the ionization level needs to be tracked in order to estimate the avalanche rate more accurately. The seed runaway density $N_r(0)$ then plays a decisive role.

In the limit of a fully ionized plasma, the avalanche rate is given by Eq.~(\ref{gamma}). Substituting Eq.~(\ref{E}) for electric field in Eq.~(\ref{gamma}) and integrating over time we see that the final number of runaway electrons does not depend on the quench time, nor on any other aspect of the quench history if $E/E_c \gg 1$ \cite{Rosenbluth, Breizman},
\begin{equation}
   \frac{N_\mathrm{r}(\infty)}{N_\mathrm{r}(0)} =  \exp \left( \sqrt{\frac{\epsilon \pi}{3 (Z+5)}} \frac{e |\chi_0|}{m_e c R \ln \Lambda}Q \right),
   \label{int_gamma}
\end{equation}
where for exponential $E$-field decay
    $$ Q = 1- \frac{E_c}{E_0}  \left( 1 - \ln{\frac{E_c}{E_0}} \right) \simeq 1$$
if $E/E_c \gg 1$. This estimate gives ${N_\mathrm{r}(\infty)}/{N_\mathrm{r}(0)} \sim 1$, which implies that no significant runaway population is generated if a quench occurs during plasma operation in W7-X.

Finally, we note that no avalanche at all is possible if $E_0 <
E_\mathrm{c}$, or
\begin{equation}
E_c t_{\rm quench} > {|\chi_0|}/{R}.
\label{avalanche criterion}
\end{equation}
For W7-X, this condition translates to 
    $$\frac{n_e}{10^{20} {\rm m}^{-3}} \cdot \frac{t_{\rm quench}}{1 \rm s}>1.$$
Runaway avalanches are thus in principle impossible in plasmas with high enough density. 

\section{Conclusions and outlook}

Runaway electrons can be generated in a stellarator if the currents in the coils are ramped down sufficiently quickly, even if there is no net toroidal current in the coils or the plasma. Although particles circulating around the torus are generally well confined in stellarators, during coil-current ramp-down electrons drift radially outward as the magnetic field decays, ultimately producing hard X-rays upon wall impact and perhaps damaging the wall. 

In W7-X, the avalanche amplification factor is of order unity during routine plasma operation, implying negligible secondary multiplication and thus little hazard from runaways in this regime. In contrast, if a quench would occur between discharges, when the gas pressure in the vacuum vessel is low, the avalanche factor can be very large, so the final runaway population depends critically on the initial seed $N_r(0)$.
In principle, no significant seed is expected in the absence of radiation-related sources between discharges,  but the relatively large potential avalanche gain makes it difficult to entirely rule out substantial runaway production. The existence of a "weak" seed — involving only a few fast electrons — remains an open question. A practical mitigation strategy could be to maintain a sufficiently high neutral gas pressure (in W7-X above $2\cdot10^{-6}$ mbar), such that $e E_0 < F_s$ for electrons with energies below 100 eV. Since the spontaneous appearance of higher-energy electrons is unlikely, this pressure threshold should effectively suppress runaway initiation.

The situation is markedly different in a reactor-scale stellarator, where:

\begin{itemize}
    \item[(a)] the ratio $\chi_0/R$ is expected to be an order of magnitude larger than in W7-X \cite{Schauer}. For example, in a reactor-size device \cite{Hegna_2025} with $R \approx 12.5 \,\mathrm{m}$, minor radius $a \approx 1.25\,\mathrm{m}$, and $B \approx 9\,\mathrm{T}$, one can expect $\chi_0/R \approx 0.5 \mathrm{T\,m}$. The conservative estimate in Eq.~(\ref{Gamma_comb}) then yields ${N_{\mathrm r}(\infty)}/{N_{\mathrm r}(0)} \sim 10^{25}$ for inter-discharge quenches at a gas density of $10^{16}\,\mathrm{m^{-3}}$ and quench duration $t_{\mathrm{quench}}$ in the range $3\text{–}10\,\mathrm{s}$;
    
    \item[(b)] with higher $\chi_0/R$, a quench during low-density operation produces substantial avalanche amplification. Indeed, Eq.~(\ref{int_gamma}) gives ${N_{\mathrm r}(\infty)}/{N_{\mathrm r}(0)} \sim 10^{5}$ for $n_e = 10^{19}\,\mathrm{m^{-3}}$. Note that Dreicer primary generation can be efficient in these conditions: it becomes significant when $E/E_D > 0.02$, where the Dreicer field is $E_D \equiv E_c\, m_e c^2/T_e$. For $t_{\mathrm{quench}} = 10\,\mathrm{s}$, giving $E_0 \approx 0.05\,\mathrm{V\,m^{-1}}$, this criterion is met, e.g. for a $T_e \sim 1\,\mathrm{keV}$ plasma;
    
    \item[(c)] in addition to the Dreicer primary generation mechanism, sources such as Compton scattering of $\gamma$-radiation from activated walls and tritium decay within the plasma provide a seed population during and between discharges.
\end{itemize}

Under these circumstances, substantial runaway generation becomes likely if the coil currents are rapidly ramped down and the plasma or neutral-gas density is below a critical level. The flux of $\gamma$-rays between plasma discharges depends on the activation history and the structural composition of the vessel wall. It is expected to be a few orders of magnitude lower than that during plasma operation, where it can be estimated to be up to the level of the neutron flux, i.e. $\Gamma_\gamma \le {P}/({S\Delta E})$, where $P$ and $S$ denote the reactor's power and surface area, respectively, and $\Delta E = 17.6$ MeV. The Compton-scattering cross-section depends on the $\gamma$-ray energy but can be crudely approximated by the Thomson cross section $\sigma_\gamma = 8 \pi r_e^2/3$. The resulting seed production rate is then $\dot n_\mathrm{seed} = n_e \Gamma_\gamma \sigma_\gamma \approx 10^6 \mathrm{m^{-3}s^{-1}}$ for $P/S = 0.5\ \mathrm{MW/m^2}$ and $n_e = 10^{16} \mathrm m^{-3}$. The energies of these Compton-produced seed electrons can reach the MeV range, making them capable of runaway acceleration in any field satisfying $E > E_c$.

These considerations suggest that it would be wise to take precautions to avoid runaway-electron generation in future large stellarators. A superconducting-coil quench is a few orders of magnitude slower than a tokamak disruption, and the problem of runaway-electron mitigation should be correspondingly simpler. There is, for instance, ample time to inject material into the torus in order to stop the runaways, but some form of intervention is likely to be necessary. 

%
%

\bibliographystyle{apsrev4-1} 
\bibliography{bibl}

\appendix
\section{End Matter}
\subsection{Kinetic theory}
The drift kinetic equation for relativistic electrons is
\begin{equation*}
  \frac{\p f}{\p t} + \dot{\bi{r}} \cdot \nabla f + \dot{p_\|}\frac{\partial f}{\partial p_\|} = C(f) + S,
\end{equation*}
where $S$ encapsulates the effect of collisions at close range and the
velocity-space coordinates have been chosen to be $ \mu =
{p_\perp^2}/(2B) $ and $ p_\| = \gamma v_\| $ with
\begin{equation*}
  \gamma = \frac{1}{\sqrt{1-v^2/c^2}} = \sqrt{1 + \frac{1}{c^2} \left(2
  \mu B(\bi{r}) + p_\|^2 \right). }
\end{equation*}
The equations of motion are \cite{Cary}
\begin{eqnarray*}
  \dot{\bi{r}} = \frac{p_\| \bi{B}^\ast}{\gamma B_\|^\ast} + \frac{\bi{E}^\ast \times \bi{b}}{B_\|^\ast},\\
  \dot p_\| = \frac{q \bi{B}^\ast \cdot \bi{E}^\ast}{m B_\|^\ast},\\
  \bi{B}^\ast = \bi{B} + \frac{p_\|}{q} \nabla \times \bi{b},\\
  \bi{E}^\ast = \bi{E} - \frac{m}{q} \left( c^2 \nabla \gamma -
  p_\| \frac{\p \bi{b}}{\p t} \right),
\end{eqnarray*}
where $q=-e$ denotes the charge, and the subscript $\|$ indicates the
component in the direction of $\bi{b} = \bi{B}/B$. The kinetic energy
$\epsilon = p_\|^2/2 + \mu B$ evolves according to
\begin{equation*}
  \dot \epsilon = \frac{q p_\| E_\|}{m} + \mu \frac{\p B}{\p t},
\end{equation*}
The last term, which describes the conversion of magnetic to kinetic
energy, is relatively small,
\begin{equation*}
  \mu \frac{\p B}{\p t}  \bigg\slash \frac{q p_\| E_\|}{m} \sim
  \frac{\gamma^2 v_\perp^2}{B} \left| \nabla \times \bi{E} \right|
  \bigg\slash \frac{q\gamma v_\| E_\|}{m} \sim \frac{v_\perp
    \rho}{v_\| R} \ll 1,
\end{equation*}
where $\rho = \gamma m v_\perp/(qB)$ denotes the gyroradius. In order
to make contact with Ref.~\cite{Rosenbluth}, we change velocity-space
variables to $(p,\lambda)$, where $p^2 = 2 \epsilon$ and $\lambda = 2
\mu/p^2$. Then $\dot p = \dot \epsilon/p$ and $ \dot \lambda = - 2
\lambda \dot \epsilon / p^2$, and the drift kinetic equation becomes
\begin{eqnarray}
  \frac{\p f}{\p t} + \dot\psi \frac{\p f}{\p \psi}
  + \dot\theta \frac{\p f}{\p \theta} + \dot \varphi \frac{\p f}{\p \varphi} + \nonumber\\
 \frac{q \xi E_\|}{m} \left[ \frac{1}{p^2} \frac{\p (p^2 f)}{\p p} - \frac{2}{p} \frac{\p (\lambda f)}{\p \lambda} \right] = C(f) + S
  \label{dke}
\end{eqnarray}
where $\xi = p_\| / p = \pm \sqrt{1-\lambda B}$. 

To solve this equation, we expand the distribution function in the smallness of the gyroradius and the electric field, $f = f_0 + f_1 + \cdots$. To lowest order in $\rho/R \ll 1$, the orbits follow field lines, i.e. $\dot \psi = 0$ and 
\begin{equation*}
  \dot \theta = \frac{v_\| \bi{B} \cdot \nabla \theta}{B} =
  \frac{\iota v_\| B}{G} = \iota \dot \varphi,
\end{equation*}
and the distribution function is thus a function of $(\psi,p,
\lambda)$. For circulating particles, the parallel-streaming terms in
(\ref{dke}) are annihilated by the operator
\begin{equation*}
  \lang \frac{B}{\xi} \left( \cdots \right) \rang,
\end{equation*}
and so is the next-order radial drift, $\lang B \dot \psi / \xi \rang = 0$, 
where angular brackets denote the usual flux-surface average \cite{ROP}. The resulting equation,
\begin{eqnarray*}
  \lang \frac{B}{\xi} \rang \frac{\p f_0}{\p t} + \frac{q \lang E_\| B \rang}{m} 
  \left[ \frac{1}{p^2} \frac{\p (p^2 f_0)}{\p p} - \frac{2}{p}
  \frac{\p (\lambda f_0)}{\p \lambda} \right] = \nonumber \\
  \lang \frac{B}{\xi} \left[ C(f_0)  + S \right] \rang
\end{eqnarray*}
coincides with that solved in Ref.~\cite{Rosenbluth}. The distribution
function and the growth rate of the avalanche in are thus essentially
the same as in a tokamak, notwithstanding non-axisymmetry and the
radial drift of the electrons due to the decaying magnetic field.

\subsection{Ionisation during initial acceleration}
We can estimate the number of secondary electrons during initial acceleration to non-relativistic energies by noting that the
expected number of ionisation events caused by an accelerating
electron is equal to
\begin{equation*}
\alpha = \int \sigma n_\mathrm{n} v \; \rmd t = \int \sigma n_\mathrm{n} v \;
\frac{\rmd w}{\dot w},
\end{equation*}
where $w$ denotes the kinetic energy of the electron, $n_\mathrm{n}$
the neutral density and $\sigma(w)$ the ionisation cross
section. Since $\dot w = eEv$, we thus have
\begin{equation}
  \alpha = \frac{n_\mathrm{n}}{eE} \int_0^{w_{\rm max}} \sigma \rmd w, \label{alpha}
\end{equation}
where the integral is taken from 0 to the maximum energy the electron reaches during the time period in question. In the non-relativistic Born approximation, 	
\begin{equation*}
\sigma \sim \frac{\sigma_0  w_0}{w}, 
\qquad w \rightarrow \infty,
\end{equation*}
where $w_0$ is some reference energy. The integral therefore diverges
at as $w_{\rm max} \to \infty$, but only logarithmically. If $w_{\rm
  max} / w_0 \gg 1$ the main contribution comes from high energies and we thus obtain the estimate
\begin{equation*}
\alpha \simeq \frac{n_\mathrm{n} \sigma_0 w_0}{eE} \; \ln \frac{w_{\rm max}}{w_0}.
\end{equation*}
Between discharges in W7-X, the pressure in the plasma vacuum vessel is about
$10^{-7}$ mbar $ = 10^{-5}$ Pa, and the hydrogen molecule density is
thus $ n_\mathrm{n} \sim 2.4 \cdot 10^{15} \; {\rm m}^{-3}$. The
ionisation cross section peaks at about 70 eV, where it reaches
$\sigma_{\rm max} \simeq 10^{-20} \; {\rm m}^2$, and has fallen to
about a quarter of this value at 1 keV. It thus appears that the
non-relativistic ionisation probability (the expected number of
ionisation events caused by an electron as it is accelerated to
sub-relativistic energies) is only of the order of
1-10\%. Non-relativistic electrons are thus not expected to produce
breakdown.

\end{document}